\documentstyle[12pt]{article}
\def\re{\par\noindent\parbox[t]{3mm}{\ }\hangindent=1cm\hangafter=1}

\newcommand{\bm}[1]{\mbox{\boldmath $ #1 $}}
\newcommand{\hs}[2]{#1 _{\mbox{\scriptsize{#2}}}}

\newcommand{\Kep}{\hs{t}{K}}

\def\ga{\hspace{1ex} ^{>} \hspace{-2.5mm}_{\sim} \hspace{1ex}}
\def\la{\hspace{1ex} ^{<} \hspace{-2.5mm}_{\sim} \hspace{1ex}}

\newcommand{\RMS}[1]{$\langle #1 ^2\rangle ^{1/2}$}
\newcommand{\RMSn}[1]{$\langle #1 ^2\rangle ^{1/2}/\hs{r}{h}\Omega_0$}
\newcommand{\RMSm}[1]{\langle #1 ^2\rangle ^{1/2}}
\newcommand{\ha}[1]{$\hs{\lambda}{#1}/\hs{r}{h}$}
\newcommand{\he}[2]{$\hs{\lambda}{#1}(#2)/\hs{r}{h}$}

\newcommand{\vs}[2]{v_{\mbox{\scriptsize{#1}}, #2}}
\newcommand{\bvs}[2]{{\bm v}_{\mbox{\scriptsize{#1}}, #2}}
\newcommand{\vsd}[2]{\langle v_{\mbox{\scriptsize{#1}}, #2} ^2 \rangle ^{1/2}}
\newcommand{\vsdn}[2]{\langle v_{\mbox{\scriptsize{#1}}, #2} %
                                           ^2 \rangle ^{1/2}/\hs{r}{h}\Omega_0}
\newcommand{\refi}[6]{
\re
{\sc #1}, #3, {\em #4}, {\bf #5}, #6, #2.
}

\newcommand{\refpre}[2]{
\re
{\sc #1}, #2.
}

\newcommand{\refb}[5]{
\re
{\sc #1}, {\em #3}, #5, #4, #2.
}

\begin{document}
\setlength{\baselineskip}{24 pt}
\renewcommand{\theequation}{\arabic{equation}}

%
%

\begin{center}
{\bf{ \LARGE
Spatial Structure and Coherent Motion\\
in Dense Planetary Rings\\ 
Induced by Self-Gravitational Instability}}
\\
\vspace{80pt}
\noindent
{\large \sc Hiroshi Daisaka and Shigeru Ida}\\[8mm]
\vspace{30pt}

\noindent
{ \it Department of Earth and Planetary Sciences,
Faculty of Science\\
Tokyo Institute of Technology, Tokyo 152-8551, Japan}\\
Telephone: +81-3-5734-2243\\
Fax: +81-3-5734-3538\\
E-mail: hdaisaka@geo.titech.ac.jp\\
ida@geo.titech.ac.jp
\vspace{30pt}
\noindent
\vspace{30pt}

\end{center}

\newpage
%
%
%
%
%

\begin{center}
\Large{\bf ABSTRACT}
\end{center}
 
We investigate the formation of spatial structure 
in dense, self-gravitating particle systems
such as Saturn's B-ring through local $N$-body simulations 
to clarify the intrinsic physics based on individual particle motion.
In such a system, Salo (1995) showed that 
the formation of spatial structure such as wake-like structure and 
particle grouping (clump) arises spontaneously due to 
gravitational instability and 
the radial velocity dispersion increases 
as the formation of the wake structure. 
However, intrinsic physics of the phenomena
has not been clarified.
We performed local $N$-body simulations including mutual gravitational
forces between ring particles as well as direct (inelastic) collisions
with identical (up to $N\sim40000$) particles.
In the wake structure
particles no longer move randomly but coherently.  
We found that particle motion was similar to Keplerian motion
even in the wake structure
and that the coherent motion was produced since
the particles in a clump had similar eccentricity and longitude of perihelion.
This coherent motion causes the increase and oscillation 
in the radial velocity dispersion.
The mean velocity dispersion is rather larger in a more dissipative case
with a smaller restitution coefficient and/or a larger surface density
since the coherence is stronger in the more dissipative case.
Our simulations showed that the wavelength of the wake structure 
was approximately given by the longest wavelength 
$\hs{\lambda}{cr} = 4\pi^2 G\Sigma/\kappa^2$ 
in the linear theory of axisymmetric gravitational instability in a thin disk,
where $G$, $\Sigma$, and $\kappa$ are the gravitational constant,
surface density, and a epicyclic frequency.

\newpage
%

%
%

\section{Introduction}

The observation by the Voyager 
revealed that Saturn's ring is not homogeneous 
but it has complex structures.
The ring has fine axisymmetric subrings with the width $\sim 10$ km
(e.g., Smith et al. 1982).
Furthermore, the observation of Photopolarimeter (PPS) stellar occultation 
showed the finer density structure (microstructure)
with scale down to 100m in B-ring (Esposito et al. 1983).
Several ideas have been proposed to explain why such structure is formed 
and maintained against viscous diffusion due to collisions
and gravitational scatterings of ring particles, 
but there is no satisfactory explanation.

To account for the axisymmetric subring structure, viscous instability 
was proposed by 
Lin and Bordenheimer (1981), Ward (1981), and Lukkari (1981).
If viscous instability arises,
diffusion from a less dense region is larger than that from a denser region
so that the density contrast is strengthen 
until non-linear effect becomes important.
The condition of this instability is $\partial(\nu\Sigma)/\partial\Sigma < 0$,
where $\nu$ and $\Sigma$ are kinematic viscosity and surface density of ring.
It has been investigated 
whether a ring particle system satisfies the condition or not.
However, negative results for viscous instability were reported
by both theoretical studies using Boltzmann equation (Araki and 
Tremaine 1986, Araki 1988, 1991) and local $N$-body simulations 
(Wisdom and Tremaine 1988, Salo 1991, 1992a, 
Richardson 1994).

On the other hand, 
Salo (1992b), Richardson (1994), and Salo (1995) gave important suggestions
on the microstructure observed in B-ring.
They extended the local {\it N}-body simulations developed by Wisdom and 
Tremaine (1988) to include self-gravitational forces between 
particles as well as inelastic collisions 
and used larger number of particles ($N>1000$) than
that in Wisdom and Tremaine (1988) ($N=40$).
They found the wake-like structure which may correspond to the substructure.
Salo (1995) showed that the wakes are created if both
self-gravity and inelastic collisions are included.
Only self-gravity or only inelastic collision does not create
such structure.
The wake structure is time dependent and transient,
being created and destroyed on the time scale of an order of Keplerian period.
The wake structure looks like fluid turbulence. 

Salo (1995) also found that
when the wake-like structure is created,
the equilibrium velocity dispersion of ring particles
always satisfies ${\cal Q} \sim 2$, where ${\cal Q}$ is
Toomre's non-dimensional parameter (Toomre 1964) defined by
\begin{equation}
{\cal Q} = \frac{\kappa c_r}{3.36G\Sigma},
\label{eq:ToomreQ}
\end{equation}
where $G$ is the gravitational constant, 
$\kappa$ is a epicyclic frequency, 
and $ c_r $ is radial velocity dispersion.

Since the wake structure does not develop when ${\cal Q} > 2$,
self-gravitational instability would be responsible for
the wake structure.
The linear calculation shows that a self-gravitating, 
differentially rotating disk of collisionless particles
becomes unstable against axisymmetric perturbation
for ${\cal Q} < 1$
(e.g., Toomre 1964, Julian and Toomre 1966).
For non-axisymmetric perturbation, 
Griv (1998) derived a similar criterion,
${\cal Q} \sim 2\Omega_0/\kappa$, where $\Omega_0$ is the angular velocity.
Griv (1998) also confirmed the above criterion by local $N$-body simulations.

Since the relation of ${\cal Q} \sim 2$ is maintained
once the wake-like structure is formed (Salo 1995),
the radial velocity dispersion $c_r$ increases with $\Sigma$.
This increase in the radial velocity dispersion must be
closely related with the formation of the wake structure.
Salo (1995) suggested that such an increase of the velocity dispersion 
would come from scatterings by the collective wakes. 
He also suggested that the systematic motion of particles
in the wakes may be responsible for the increase 
in the radial velocity dispersion.
However, it has not been clarified how the radial velocity increase
is related with the formation of the wake structure.

The study on the wake structure is important to 
clarify the microstructure
in Saturn's B-ring.
Furthermore, it is also important in the issue of the stability 
of Saturn's B-ring,
since the transient wake structure would induce relatively
large angular momentum flux to quickly diffuse out the ring.
(The detailed analysis of the angular momentum transfer will
be done in the next paper).

We perform local $N$-body simulations including both mutual 
gravitational forces between particles and inelastic collisions as 
Salo (1995) did.
We focus on the problem of 
the relation between the wake formation and the velocity
increase through detailed analysis
about how self-gravitating particles behave in the wake-like structure.

Simulation method is described in section 2.
In section 3, we first compare our results with previous works 
(especially with Salo 1995),
then we will analyze the coherent particle motion in more detail.
We find regular oscillation of the velocity dispersion 
associated with the increase in the time averaged velocity dispersion
and clarify the relationship between the wake structure and the
velocity change.
We also analyze the scale of the structure by performing simulations 
with up to $N \sim 40000$ particles.
In section 4 we summarize the results and give discussion.

\newpage
%
%

%
%

\section{Numerical methods}

\subsection{Model description}

We adopt local {\it N}-body method which was first applied for 
the study of a dense ring system by Wisdom and Tremaine (1988)
and followed by Salo (1991, 1992a, 1992b, 1995) and Richardson (1994).
The ``local'' means that we consider a box with width $L_x$ and height $L_y$
at a semimajor axis $a_0$ in the ring, which revolves in a circular
orbit with the Kepler angular velocity $\Omega_0$ 
at the reference point $a_0$ and 
is small compared to the width of the whole ring (see Fig. 1). 
Motion of particles is pursued only in this box with 
the periodic boundary conditions.
This method would be valid because we are considering the structure with
much smaller scale than the width of the B-ring and the orbits of ring 
particles are nearly circular in the coplanar plane.

We use the Cartesian coordinates at $a_0$ which rotate with 
the angular velocity $\Omega_0$, 
in which $x$-axis points radially outward, 
$y$-axis points to the direction of orbital motion of the coordinate origin, 
and $z$-axis points to the direction normal to the orbital plane.
Motion of a ring particle $i$ is described by Hill's equation
(e.g., Hill 1878, Nakazawa and Ida 1988):
\begin{equation}
\begin{array}{@{\,}ccccccc}
\ddot{x} _i & = & 2 \Omega _0 \dot{y} _i & + & 3 \Omega _0 ^2 x_i & + & 
\sum ^N _{j \neq i} \displaystyle\frac{Gm_j}{r_{ij}^3} (x_j - x_i), \\
\ddot{y} _i & = & -2\Omega _0 \dot{x} _i &   &                    & + &
\sum ^N _{j \neq i} \displaystyle\frac{Gm_j}{r_{ij}^3} (y_j - y_i), \\
\ddot{z} _i & = &                        & - & \Omega _0 ^2 z_i   & + & 
\sum ^N _{j \neq i} \displaystyle\frac{Gm_j}{r_{ij}^3} (z_j - z_i),
\end{array}
\label{eq:hill}
\end{equation}
where $ m_j $ and $r_{ij}$ are mass of particle {\it j}
and the relative distance between particle {\it i} and {\it j},
and $\Omega_0 = \sqrt{G\hs{M}{s}/a_0^3}$,
$ \hs{M}{s} $ is mass of a central planet.
The first and second terms on the right hand sides of Eqs.~(\ref{eq:hill}) 
denote the Coriolis and the tidal force, 
and the last terms are the mutual gravitational force between ring particles.
When the mutual gravitational force can be neglected, 
Eqs.~(\ref{eq:hill}) are solved analytically and the solution represents
an orbit of the Keplerian motion (see Nakazawa and Ida 1988).

Scaling and parameters which 
characterize a ring system are as follow.
Throughout the present paper,
we assume that 
all particles have the same radius $\hs{r}{p}$ and mass $m$.
The mutual Hill's (tidal) radius is defined by 
$ \hs{r}{h} = h a_0 $, where
$ h $ is the reduced Hill's radius defined as
\begin{equation}
h = (\frac{2m}{3\hs{M}{s}})^{1/3}.
\end{equation}
$r_h$ represents the radius of potential wall between two particles
at which central force (tidal force) and mutual gravitational force between
the particles are balanced.
Within the radius, mutual gravity dominates central force and vise verse.
It is possible to rewrite Eqs.~(\ref{eq:hill}) as mass-independent forms
by scaling length, time, and mass as
\begin{equation}
	\begin{array}{@{\,}ccl}
	\tilde{x} & =  & x/\hs{r}{h}, \\
	\tilde{t} & = & t\Omega_0, \\
	\tilde{m} & = & m/\hs{M}{s} h^3=\displaystyle\frac{3}{2}.
	\end{array}
\end{equation}
The scaled mass $\tilde{m}$ is independent of mass of particle.
In our simulations, we use these scaled variables and solve 
non-dimensional equations of Eqs.~(\ref{eq:hill}).

A ring system is characterized by two non-dimensional parameters,
optical depth $\tau$ and the ratio $\hs{r}{h}/2\hs{r}{p}$.
The dynamical optical depth is defined as
\begin{equation}
\tau= \frac{N \pi \hs{r}{p} ^2}{L_x L_y}
\end{equation}
where $ L_x \times L_y $ and $N$ are the domain area
and total number of particles in the domain.
The optical depth can be expressed as $\tau \sim \hs{t}{K}/\hs{t}{c}$ 
(e.g., Goldreich and Tremaine 1982), 
where $\hs{t}{K}$ and 
$\hs{t}{c}$ are the Keplerian time and the mean collision time.
If $\tau$ is large enough (typically, $\tau \ga 1$), 
clumps tend to be formed due to collision damping and self-gravity,
while the clump formation is inhibited by differential rotation
of the Keplerian motion if $\tau<1$, as shown later.

The other parameter $\hs{r}{h}/2\hs{r}{p}$ is 
\begin{equation}
\frac{\hs{r}{h}}{2\hs{r}{p}} \simeq 
        0.82 (\frac{\rho}{900\mbox{kg/m}^3})^{1/3}
                (\frac{a_0}{10^{8}\mbox{m}}). 
\label{eq:rhrp}
\end{equation}
where Saturn's mass, $\hs{M}{s} = 5.69 \times 10^{26}$kg, is assumed
and $\rho$ represents material density of a ring particle.
Since $\pi(2r_p)^2$ and $\pi r_h^2$ express geometrical and 
characteristic gravitational cross sections for small random velocity,
this parameter regulates which mechanism dominates velocity
changes, gravitational scatterings or direct collisions.
If $\hs{r}{h}/2\hs{r}{p} \ga 1$, gravitational scatterings is 
more effective than direct collisions and vice versa
(Ohtsuki 1999). 
For the B-ring, 
$\hs{r}{h}/2\hs{r}{p} \sim 1$, so that both gravitational scatterings
and direct collisions are important.
On the other hand,
ratio $ \hs{r}{h}/2\hs{r}{p} $ is also related to 
the efficiency of the self-gravity against the tidal force of 
the central planet.
Ohtsuki (1993) showed that the tidal force inhibits formation of
persisting aggregates bounded by self-gravity if $\hs{r}{h}/2\hs{r}{p} < 1$.
From Eq.~(\ref{eq:rhrp}), the condition $\hs{r}{h}/2\hs{r}{p} > 1$
holds outside A-ring.
The systems with the same 
$\hs{r}{h}/2\hs{r}{p}$ have the same dynamical condition, 
even if $a_0$ and $\rho$ are different.
We examine the dependence of the wake and clump formations on 
$\hs{r}{h}/2\hs{r}{p}$ rather than on $a_0$ and $\rho$.

In our simulations, we set up initial conditions as follows.
The positions of particles are randomly
distributed in the simulation region, avoiding overlapping of particles.
The velocities except for the shear velocity of 
individual particles $3\Omega_0 x_i/2$ are 
chosen randomly in the limited range from $0$ to $5 \hs{r}{h}\Omega_0$, but
when the initial random velocity yields $\hs{{\cal Q}}{initial} < 2$
we adopt random velocity large enough to be $\hs{{\cal Q}}{initial} > 2$.
The density structure obtained by our simulations is independent of choice of 
random numbers used to generate these initial conditions.
The initial velocity is immediately relaxed and the equilibrium velocity state 
is established within a few Keplerian periods.

Equations~(\ref{eq:hill}) is integrated with the fourth-order Hermite scheme, 
which is one of predictor-corrector integrators and needs
time derivatives of the mutual gravitational force 
as well as the mutual gravitational forces
instead of past few data of position and velocity
for interpolation (Makino and Aarseth 1992).
This integrator is easily implemented to the simulations in which 
there are discontinuous phenomenon such as 
direct collisions and jumps due to the periodic boundary conditions,
because the Hermite integrator does not need data of past steps.
In the integration, most expensive part of integration is the calculation of 
the mutual gravitational forces and its time derivatives
whose calculation cost is $O(N^2)$.
To reduce computational time,
mutual force and its time derivative are calculated by HARP-2, 
which is a special purpose hardware for calculating gravitational force
(Makino et al. 1993). 
``HARP'' means Hermite AcceleratoR Pipeline and is one kind of
GRAPE system (``GRAPE'' means GRAvity PipE; Sugimoto et al. 1990).
This hardware is connected to a host computer 
through a communication interface.
The host sends position, velocity, and mass of all particles to the hardware.
The hardware calculates gravitational forces with pipelines and 
returns them to the host.
The host integrates orbits utilizing the returned gravitational forces.
By using this hardware, calculation cost is significantly reduced.

\newpage

%
%

%
%

\subsection{Calculation of mutual gravitational force and direct collision}

As stated above,
the mutual gravitational force between ring particles and the direct 
(inelastic) collision play an important role in the evolution and 
the kinetic behavior of particle system, 
and the wake and clump formations.
An equilibrium state is established by the balance between viscous energy
gain (heating) through
gravitational scatterings and direct collisions and energy dissipation 
(cooling) through inelastic collision.

In the original Wisdom and Tremaine's method,
the mutual gravity was ignored or was only treated as vertical force by 
enhancement of vertical frequency.
Salo (1992b, 1995) and Richardson (1994)
extended simulations exactly including mutual force.
Richardson (1994) used the tree method while
Salo (1995) carried out direct sum of forces exerted on a
particle from neighbor particles within the distance $\hs{R}{max}$.
The cut-off length $\hs{R}{max}$ is often substantially less than width of 
simulation region.
In our simulations, 
we adopt direct sum of forces from all particles 
with some adjustment explained below.

Salo (1995) studied the influence of cut-off distance $\hs{R}{max}$ 
and found an optimal value.
But the larger $\hs{R}{max}$ may be needed when the strong wake-like structure 
appears and coherent motion in the structure is important.
We consider gravitational forces exerted on a particle from all 
other particles used in the simulation,
carefully treating particles near boundary as well as in Salo (1995).
For the effective use of HARP,
we introduce the following subregion method.
As in Fig. 1, we consider an original region (a box with thick line) 
and its copies.
Because of the Keplerian shear velocity, 
radially inner and outer boxes have different angular velocity so that 
these boxes slide upward and downward with the shear velocity
$3\Omega_0L_x/2$.
A particle at $ (x,y,z) $ in the central box with the velocity 
$ (v_x,v_y,v_z) $
has its images at $ (x \pm nL_x, y \pm mL_y \mp 3\Omega_0 n L_x t/2, z) $ with 
the velocity $ (v_x,v_y \mp 3\Omega_0 n L_x/2,v_z )$, 
where $n$ and $m$ are integers.
When a particle in the central box outgoes from a certain boundary, 
its image at the neighbor box comes from the opposite boundary.
We divide the simulation region into nine subregions, represented as broken
line in Fig. 1. 
For each subregion, we make virtual regions with the same size of original.
In the virtual regions, the subregion is centered.
Dark and light shaded regions in Fig. 1 are the example of 
a subregion and its virtual region.
For particles in the subregion, we count gravitational forces of all
particles in its virtual region.
In this method, it happens that force from closer particle 
is neglected but that from more distant particle is counted.
However, this asymmetry occurs in the in the outer part of the virtual region 
and does not affect the results 
because the size of the simulation is considerably larger than $\hs{R}{max}$
Salo (1995) used.
Actually we confirmed it performing the simulations with different size of 
simulation region.

In our simulations, we adopt a hard-sphere collision model 
that is commonly used in previous simulations
(Wisdom and Tremaine 1988, Salo 1991, 1992a, 1992b, 1995, Richardson 1994).
A collision changes only impact velocity in normal direction.
Normal and tangential components of relative velocities of 
colliding particles after the collision, $v_n'$ and $v_t'$, are described 
by 
\begin{equation}
\begin{array}{@{\,}ccl}
v_{n}' & = & -\epsilon v_{n}, \\ 
v_{t}' & = & v_{t}
\end{array}
\label{eq:collision}
\end{equation}
where $ \epsilon $ is the restitution coefficient of ring particles.
Generally, $ \epsilon $ depends on its material properties as well as
the impact velocity.
In spite of many efforts (Bridges et al. 1984, Hatzes et al. 1988,
Dilley 1993, Supulver et al. 1995), 
we have had no precise knowledge on $ \epsilon $ yet.
In this study, we will treat $ \epsilon $ as a parameter.

Detection of collision is one of difficulties in treating collisions.
Ideally, collisions should be detected at the instance when colliding pairs
just touch each other but it is practically impossible.
Such problems were discussed by Richardson (1994) and Salo (1995).
In this study, we use a simple method for detection of collision.
Collision is detected as overlapping of colliding particles 
approaching each other.
We search colliding pairs by using a neighbor list 
of each particle which is calculated by HARP-2 as well as
mutual forces and its time derivatives.
If a pair in the neighbor list overlaps,
$ r_{ij} \leq r_{\mbox{\scriptsize{p}}, i} + r_{\mbox{\scriptsize{p}}, j} $,
and is approaching,
$ \bm{r}_{ij} \cdot \bm{v}_{ij} <0 $, 
where $ \bm{r}_{ij} $ and $ \bm{v}_{ij} $
are relative distance $ \bm{r}_{ij} = \bm{r}_{j} - \bm{r}_{i} $
and velocity $ \bm{v}_{ij} = \bm{v}_{j} - \bm{v}_{i} $ and 
$ r_{\mbox{\scriptsize{p}}, i} $
is radius of particle $i$, 
we assume that these particles collide
and velocity of colliding pair after the
collision is calculated by Eqs.~(\ref{eq:collision}).
If $ r_{ij} \leq r_{\mbox{\scriptsize{p}}, i} + r_{\mbox{\scriptsize{p}}, j} $
but $ \bm{r}_{ij} \cdot \bm{v}_{ij} > 0 $, 
we let the particles separate.
However, colliding bodies can happen to overlap significantly 
without having been detected.
In this case, the bodies keep sinking,
which stops calculation (Richardson 1994, Salo 1995).
To avoid this trouble, 
we remove overlapping of colliding bodies by the method 
used by Richardson (1994).
In his method, overlapping bodies are moved outward along the line 
connecting each center to the location 
where these particles are just touching. 
From conservation of center of mass of colliding particles,
adjusted positions are given as
\begin{equation}
\begin{array}{@{\,}ccl}
\bm{x}_i ' & = & \bm{x}_i  - \Delta \bm{r},\\
\bm{x}_j ' & = & \bm{x}_j  + \Delta \bm{r},
\end{array}
\end{equation}
where the prime denotes position after correction and
\begin{equation}
\Delta \bm{r} = 
	\frac{r_{\mbox{\scriptsize{p}},i} 
		+ r_{\mbox{\scriptsize{p}},j} - r_{ij}} {2r_{ij}} \bm{r}_{ij}.
\end{equation}
We always apply this adjustment when collision is detected.

The penetration depth of colliding bodies can be controlled by changing
accuracy of timestep.
In our simulations, we adopted shared, variable timestep because of
frequent direct collisions and mutual gravitational forces.
The formula of timestep is simply given by 
\begin{equation}
\delta t = \min _{i} ( \alpha \frac{|\bm{a}_i|} {|{\dot{\bm{a}_i}}|} )
\end{equation}
where $\alpha$ is timestep coefficient and $ \bm{a}_{i} $ 
and $ \bm{\dot{a}}_i $ are acceleration of particle $i$ and 
its time derivative.

\newpage

%
%
%

\section{Results of simulations}

\subsection{Comparison with Salo(1995)}

In this study, we restrict our simulations to identical particle systems
in order to understand the basic dynamical properties.
Previous simulations performed by Salo (1992b) and Richardson (1994)
included the effect of a mass (size) distribution of ring particles
so that our simulation results can not be compared directly with their results.
On the other hand, Salo (1995) performed simulations of identical
particle systems with a wider range of parameters and 
obtained statistical quantities such as radial velocity dispersion.
First we perform simulations with the same parameters as used in Salo (1995)
and confirm his results.

Salo (1995) found the spontaneous formation of the wake structure
in identical particles
if self-gravity of ring particles is effective as well as 
inelastic collision,
while previous studies without the self-gravity by Wisdom and Tremaine (1988) 
and Salo (1991, 1992a) did not find such a structure.
The structure is non-axisymmetric one such as 
collective Julian-Toomre wakes (Julian and Toomre 1966).
The formation of the wake structure should be related with gravitational 
instability.
Salo (1995) performed simulations with various ring parameters such as 
$\tau$, $\epsilon$, $a_0$, 
and $\rho$ (last two correspond to $\hs{r}{h}/2\hs{r}{p}$) and
showed that the formation of the wake structure is regulated by the condition
of the Toomre parameter $\cal{Q}$ value 
for gravitational instability (Eq.~(\ref{eq:ToomreQ})).
Small $\epsilon$ and large $\tau$ lead to gravitational instability,
because small $\epsilon$ results in small velocity dispersion and
large $\tau$ results in large surface density.
For small $\hs{r}{h}/2\hs{r}{p}$, formation of persistent aggregates 
by accretion are suppressed by the tidal force.

Formation of spatial structure influences the motion of individual particles.
Salo (1995) showed that as the wake structure grows, 
the radial velocity dispersion increases with a large magnitude of 
its fluctuation and tends to keep the relation ${\cal Q}\simeq2$.
Salo (1995) also discussed the equilibrium radial velocity dispersion.
He showed that when $\hs{r}{h}/2\hs{r}{p}$ is small, i.e., 
the self-gravity is insignificant, 
the velocity dispersion is dominated by direct collisions 
rather than the self-gravity
to have a value of an order $\sim \hs{r}{p} \Omega$,
but when $\hs{r}{h}/2\hs{r}{p}$ is large but the surface density is 
small enough to avoid gravitational instability, 
gravitational encounters dominate and the velocity dispersion attains 
the surface escape velocity
$\hs{v}{es} = \sqrt{2Gm/\hs{r}{p}}$,
where $m$ is mass of a particle.
Thus, equilibrium velocity may be altered by the parameter which characterizes
a ring system as well as inelasticity of a ring particle.

Our simulations show quite similar results to Salo (1995).
Figure 2 is the spatial distribution of particles 
viewed face on and edge on.
In Fig. 2a and 2b, the optical depth $\tau$ is changed with fixed, 
velocity-independent coefficient restitution $\epsilon=0.5$ and 
$\epsilon$ is changed with fixed $\tau=0.4$, respectively.
The other parameters are the same in all simulations;
$ \hs{r}{h}/2\hs{r}{p} = 0.82 $ corresponding to, for example, 
the situation with internal density 
$\rho_0 = 900$kg/m$^3$ and Saturncentric distance $a_0 = 1.0 \times 10^{8}$m,
and $ \hs{r}{p} = 1$m.
The simulation region is square with the width $L=112$m 
($L=68.3\hs{r}{h}$).
The optical depth is controlled by changing number of particles, for examples, 
we use $N=400$ for $\tau=0.1$ and $N=2400$ for $\tau=0.6$, respectively.
Fig. 2c is the results of the simulations with various 
ratio $\hs{r}{h}/2\hs{r}{p}$ 
for $\tau=0.4$ and $\epsilon=0.5$.
In these simulations, the region is also square and
the width of the region and number of particles are
$L=59.3\hs{r}{h}$ ($L=80$m) and $N=810$ for $\hs{r}{h}/2\hs{r}{p}=0.675$,
$L=68.3\hs{r}{h}$ ($L=112$m) and $N=1600$ for $\hs{r}{h}/2\hs{r}{p}=0.82$, and
$L=70.0\hs{r}{h}$ ($L=140$m) and $N=2496$ for $\hs{r}{h}/2\hs{r}{p}=1.00$, 
respectively.
These figures correspond to Fig. 10b, 11c, and 9c (and Fig. 14) 
in Salo (1995), respectively.
Figure 2 shows that the increase of $\tau$ and/or 
ratio $\hs{r}{h}/2\hs{r}{p}$, 
and the decrease of $\epsilon$ lead to the formation of 
the strong wake structure, as pointed out by Salo (1995).
This wake-like structure is formed after a few Keplerian periods
from the beginning of the simulation.
The wake structure changes with time.
It is formed and dissolved successively on 
time scale of an order of a Keplerian period, 
so that the evolution of spatial structure affects the equilibrium velocity 
dispersion.

Figure 3 shows the equilibrium velocity of $ \langle v_x^2 \rangle ^{1/2} $ 
and $ \langle v_z^2 \rangle ^{1/2} $ as functions of $\tau$ and $\epsilon$
for the simulations of Fig. 2a and 2b.
In Fig. 3b, we also show the results with $\tau=0.1$.
Each equilibrium value and error bar are the time average and the fluctuation 
of each velocity component after initial relaxation (see Fig. 4).
In simulations of Fig. 2 and Fig. 3, the initial relaxation time is 
a few Keplerian periods. 
As shown in Fig. 4, the dispersion \RMS{v_x} in the structured case 
($\tau=1.2$) oscillates regularly with time.
In this case, the bars represent the amplitude of the oscillation.
The dotted and dashed lines in Fig.3a are 
the critical velocity calculated 
from Toomre's $\cal{Q}$ being ${\cal Q}=1$ and ${\cal Q}=2$, respectively.
Table I and II list the equilibrium values of $\langle v_x^2 \rangle ^{1/2}$ 
scaled by $\hs{r}{h}\Omega_0$ obtained by Salo's and our simulations 
with various $\tau$ and $\epsilon$.
In Fig. 3, filled circles and open triangles are our results and Salo's,
respectively.
They are in good agreement with each other.

From the comparison of Fig. 3 with Fig. 2, 
it is evident that the formation of the wake structure 
leads to the large increase in the equilibrium radial velocity 
dispersion with the large amplitude of its fluctuation, as pointed out by 
Salo (1995).
On the other hand, any drastic change 
is not observed in vertical velocity dispersion \RMS{v_z}
even if a strong wake structure is formed.
Figure 3a shows that \RMS{v_x} increases with $\tau$, 
following the line of $ {\cal Q}=2 $.
This tendency of \RMS{v_x} to follow
$ {\cal Q} \simeq 2 $ is commonly observed in simulations with
other parameters which allow the formation of the wake-like structure.
Figure 3b shows that in the region $\epsilon \ga 0.6$ for $\tau=0.4$ and
$\epsilon \ga 0.4$ for $\tau=0.1$, 
\RMS{v_x} is reduced as decreasing $ \epsilon $ 
because the collisional dissipation becomes strong.
Thickness of ring is also reduced as clearly seen in $x$-$z$ plot of 
particle distribution for $ \epsilon = 0.6 $ and $ 0.7 $ in Fig. 2b.
On the contrary, in the region $ \epsilon \la 0.6 $ for $\tau=0.4$,
\RMS{v_x} increases with decrease in $\epsilon$.
In the case of $\tau=0.1$, $\tau$ is too small to cause
gravitational instability,
so that the increase of \RMS{v_x} is not found.
In this case, gravitational scatterings dominate and the equilibrium velocity 
is given by the surface escape velocity, as pointed out by Salo (1995).
This tendency is also observed in Fig. 3a at small $\tau$. 
Note that in the case where $\hs{r}{h}/2\hs{r}{p}$ is small, 
the equilibrium velocity should be dominated by 
direct collision rather than by gravitational scattering.
Similar results were obtained by Ohtsuki (1999), who investigated 
the equilibrium velocity dispersion in the dilute ring system 
($\tau \ll 1$)
by numerical three-body integration of particle's orbits taking into account 
finite size of particles.
When $\epsilon$ is sufficiently large ($\epsilon > 0.7$ for $\tau=0.4$,
$\epsilon>0.6$ for $\tau=0.1$),
collisional damping cannot equilibrate with stirring due to 
gravitational scatterings and direct collisions, so that
both \RMS{v_x} and \RMS{v_z} monotonically increase.
These results approximately agrees with the critical restitution coefficient 
for the existence of a steady state obtained by Goldreich and Tremaine (1978) 
and Ohtsuki (1992, 1999).

Thus, our simulations as well as Salo (1995) show that the formation of 
the wake structure occurs in a dense ring system with 
self-gravitating particles and 
that as the spatial structure develops, the radial velocity 
dispersion increases with a large magnitude of fluctuation and
takes ${\cal Q} \simeq 2$.
Salo (1995) suggested that such increase of the velocity dispersion is 
due to scatterings by collective wakes. 
However, he did not give detailed argument.
On the other hand, Salo (1995) suggested through the velocity field of 
particles in the wake that the velocity increase comes from the 
difference of systematic motions between adjacent wakes.
Next we will analyze the motion of particles in detail 
to clarify the relation between the velocity increase and the wake structure.

\newpage

%

%
%

\subsection{Coherence of particle's motion in the wake structure}

As shown in the preceding subsection,
the formation and the evolution of the spatial structure and 
the motion of particles influence each other.
To clarify the relation, we analyze motion of particles in 
the wake-like structure in detail.

First we compare the evolution of the radial velocity dispersion 
in the cases with and without wake-like structure.
We perform two large $N$-body simulations for $\tau=0.3$ and 
$\tau=1.2$, which correspond to the non-structured and the structured cases,
respectively (see spatial distributions in Fig. 4).
The number of particles and the width of the square simulation region are 
$N=5052$ and
$L=230$m ($L=170.5 \hs{r}{h}, L/\hs{\lambda}{cr}=16.54$) for $\tau=0.3$, and
$L=115$m ($L=85.2 \hs{r}{h},  L/\hs{\lambda}{cr}=2.07$) for $\tau=1.2$.
The other parameters are fixed as $\epsilon = 0.4$ and 
$\hs{r}{h}/2\hs{r}{p} = 0.675$.
Figure 4 shows the evolution curves of the velocity dispersion \RMS{v_x} and 
the spatial distributions in two cases.
For $ \tau=0.3 $, the wake structure is very weak 
because of low surface number density and \RMS{v_x} 
attains a steady state with a small magnitude of fluctuation after a few 
Keplerian periods.
On the other hand, for $\tau=1.2$, the velocity dispersion increases with
oscillation with a large amplitude after initial relaxation time 
($\sim 1 \Kep$).
In the beginning of the simulation, \RMS{v_x} is rapidly reduced 
by the dissipation of random motion energy through inelastic collisions, 
and it becomes small $ \sim 2.5 \hs{r}{h}\Omega_0$.
But, after $t = 1 \Kep$, \RMS{v_x} begins to increase, 
accompanied by oscillation.
This velocity increase is closely associated with the formation of
the strong wake structure.

Similar results are also observed in simulations with different sets of 
parameters in which wake structure is formed.
To examine the effects of the boundary conditions,
we perform simulations with various sizes of simulation domain 
in the structured case,
fixing $\tau=1.2$, $\epsilon=0.4$, and $\hs{r}{h}/2\hs{r}{p}=0.675$
(see table III).
Figure 5 shows the equilibrium velocities as a function of 
number of particles $N$ (equivalently, the size of the simulation domain).
Basic features are the same in all simulations,
although it seems that \RMS{v_x} increases slightly with 
the increase of $N$ (simulation domain).
The time evolution curves of \RMS{v_x} also show oscillation with 
a similar amplitude and a period.

Salo (1995) showed similar oscillation of \RMS{v_x} 
but the oscillation is not well resolved and recognized as ``fluctuation''.
The reason is that the interval of his sampling time was large 
($\sim 0.5 \Kep$ in Fig.2 in Salo 1995, 
in our simulations the interval is $0.016 \Kep$).
We measure a period of the oscillation by Fourier transform methods.
Figure 6 shows the results of this spectrum analysis 
in the non-structured and the structured cases.
We find a strong peak at $\sim 0.6 \Kep$ for $\tau=1.2$ but not for $\tau=0.3$.
Figure 7a is the results with various computational region for $\tau=1.2$
and show similar peak period. 
Figure 7b, 7c, and 7d are the results of simulations for various $\tau$,
$\epsilon$, and $\hs{r}{h}/2\hs{r}{p}$, with the other parameters fixed.
The dependence of the period on $\tau$ and $\epsilon$ is small.
But the change of ratio $\hs{r}{h}/2\hs{r}{p}$ affects 
the period of oscillation.
Figure 7d shows that the width of peak and the period become larger 
with the increase in $\hs{r}{h}/2\hs{r}{p}$.
Simulations in the structured cases always exhibit such an oscillatory feature 
and it would be related with the property of motion of particles.

We analyze the concurrent evolution of the spatial and the velocity 
distributions in detail.
Figure 8a and 8b show the time sequence of the spatial distribution of 
particles from $t = 8.91 \Kep$ to $10.19 \Kep$ and 
the corresponding velocity distribution in the case of $\tau=1.2$.

As seen in Fig. 8a, temporal clumps are formed due to
gravitational instability and dissolved due to shearing motion.
The change of spatial structure looks like turbulent fluid,
where vortices are formed and dissolved successively.
Velocity distribution as well as spatial distribution is not steady.
Figure 8b shows that the velocity distribution revolves clockwisely, 
holding bar-like configuration.
The configuration is similar at $t = 8.91$, $9.55$, and $10.19 \Kep$,
so that the period of the revolution is about $1.3 \Kep$.
The oscillation in the evolution curve of \RMS{v_x} in the structured case 
observed in Fig. 4 would come from revolution of $v_x$-distribution.
The periods obtained by the Fourier analysis of the evolution curve of 
\RMS{v_x} (Fig. 7) and 
in the velocity distribution are different by just a factor 2.
This is why \RMS{v_x} is mean square of $v_x$, i.e., 
the velocity dispersion takes maximum value twice during a period of 
the revolution of the configuration in velocity space.
On the other hand, such change in the velocity distribution
is not observed in the non-structured case.
Figure 9 shows the velocity distribution for $\tau = 0.3$ at $t = 14.3 \Kep$.
The distribution is independent of time.

In the theoretical studies (e.g. Goldreich and Tremaine 1978, Araki and
Tremaine 1986), distribution function was assumed to be Gaussian distribution.
In the optically thin case ($\tau \ll 1$),
$N$-body simulation showed that velocity distribution is Gaussian
(Ida and Makino 1992).
The cumulative distributions of $v_x$ at different three time of 
$t=9.55$, $9.71$, and $9.87 \Kep$ in the structured case with $\tau=1.2$
are shown in Fig. 8c and that at $t = 11.1$, $12.7$,
and $14.3 \Kep$ in the non-structured case with $\tau=0.3$ are shown in Fig. 9.
The velocity distribution is always given by 
Gaussian distribution in the latter case while it deviates from 
Gaussian distribution in the former case.
With the Kolmogonov-Smirnov test (see, e.g., Press et al. 1986),
we calculate the probability which represents the similarity between 
a distribution and Gaussian distribution.
The probability is 
$P=0.64$ at $t=11.1 \Kep$,
$P=0.38$ at $t=12.7 \Kep$, and
$P=0.41$ at $t=14.3 \Kep$, respectively, in the latter case.
If $P=1$, two distributions are the same 
but if $P \ll 1$ they are different.
In the structured case, 
$P=8.18 \times 10^{-13}$ at $t=9.55 \Kep$,
$P=1.52 \times 10^{-20}$ at $t=9.71 \Kep$, and 
$P=5.52 \times 10^{-70}$ at $t=9.87 \Kep$, respectively.
Thus our results show that in the non-structured case 
the velocity distribution of $v_x$ is expressed by Gaussian distribution 
with high accuracy but in the structured case 
the distribution is far from Gaussian.

How do individual particles behave in the wake-like structure?
The $x$-component of position and 
the velocity of a Keplerian particle {\it i} relative to
the local shear velocity are written as
(Nakazawa and Ida 1988)
\begin{equation}
\begin{array}{@{\,}lcl}
x_i       & = & a_0 b_i - a_0 e_i \cos(\Omega_0 t-\delta_i),      \\
v_{x,i}   & = & a_0 \Omega_0 e_i \sin(\Omega_0 t-\delta_i),       \\
v_{y,r,i} & = & \displaystyle\frac{1}{2} a_0 \displaystyle\Omega_0 e_i 
					\cos(\Omega_0 t-\delta_i),
\end{array}
\label{eq:hill-kep-v}
\end{equation}
where $e_i$ and $ \delta_i$ are the eccentricity and 
the longitude of perihelion,
and $a_0(1+b_i)$ is the semimajor axis of particle $i$.
$v_{y,r,i}$ denotes $y$-component of velocity subtracted
the local shear velocity,
$v_{y,r,i} = v_{y,i} + 3\Omega_0x_i/2$.
If motion of particles is true Keplerian,
orbital elements such as $e_i, \delta_i$ and $ b_i $ are constants
and the evolution curve of $x$ ($v_x$) of each particle 
represents a regular oscillation in $x$ ($v_x$) with a period of $1 \Kep$ and 
with an amplitude of $e a_0$ ($ e \Omega_0 a_0$).
Mutual interactions such as direct collisions and self-gravity
should cause motion of particle to deviate from true Keplerian orbit.
Figure 8d and 8e show the time evolutions of $x$-components of 
position and velocity of arbitrary chosen three particles 
from $t=16 \Kep$ to $t=30 \Kep$ in the structured case used in Fig. 8a. 
On timescale of a few $\Kep$,
regular oscillation which comes from Keplerian epicyclic motion 
is seen in both $x$ and $v_x$ 
but the period of this oscillation is slightly larger than $1 \Kep$.
On longer timescale, the motion deviates from purely Keplerian orbit.
We call such motion as ``pseudo''-Keplerian motion.

The revolution of the velocity distribution with a period of an order of 
a Keplerian period observed in Fig. 8b would reflect 
``pseudo''- Keplerian motion of particles in the wake.
From Eqs.~(\ref{eq:hill-kep-v}),
if perturbation by other particles is negligible,
trajectory of a certain particle in velocity space is a
ellipse with 2:1 axis ratio elongated to the $v_x$-direction expressed as 
$ (v_{x,i}/a_0\Omega_0 e_i)^2 + (v_{y,r,i}/(\frac{1}{2}a_0\Omega_0 e_i))^2=1 $
and the particle revolves in the clockwise direction.
If all particles have similar phases $\delta_i$ or $\delta_i+\pi$,
phase space motion would look like that in Fig. 8b.
This means that particles move coherently in the strong wake structure,
holding the property of ``pseudo''-Keplerian motion.
Self-gravity of particles may force their coherent motion
with the mechanism proposed to maintain the elliptic ring around Uranus
proposed by Goldreich and Tremaine (1979).
If the motion is true Keplerian,
the oscillation period should be equal to $1 \Kep$.
But in the above structured case, the period observed in our simulations 
is about $1.3 \Kep$, apparently longer than $1 \Kep$.
The increase in the period may be due to friction 
caused by inelastic collisions.

Next we analyze spatial distribution of particles,
comparing it with the velocity distribution.
In Fig. 10, we mark particles in 3 local spatial regions 
by filled circles, triangles, and squares 
and also mark the corresponding particles in the velocity space,
both in non-structured and structured cases.
Spatial distributions used here are the results in Fig. 4.
In the non-structured case with $\tau=0.3$,
the marked particles are scattered randomly in the velocity space,
which means that there is no correlation of the velocity among 
neighbor particles, i.e., particles move almost randomly.
But for $\tau=1.2$, as seen in Fig. 10b, 
the neighbor particles in the real space are also localized in 
the the velocity space.
This indicates that neighbor particles have a similar velocity.
Thus, in the wake, particles no longer move randomly but coherently
(i.e., neighbors move together in the same direction 
at a similar velocity).

The coherent motion of particles is gradually destroyed 
as the neighbor particles are separated from each other by shearing motion.
We found that the marked particles diffuse in both space and velocity space 
on the time scale of an order of a Keplerian period.
But a new coherent group is created from diffused distribution 
by the gravitational instability.
Continuous creation and disruption of grouping occur in the wake.

The coherent motion affects the mean velocity of particles.
The bulk velocity due to the coherent motion increases 
the radial velocity dispersion.
Consider the population of particles marked by triangles in Fig. 10.
In the non-structured case, 
triangles are distributed randomly in the velocity space
so that the coherent motion is weak and 
the width of the distribution represents the randomness of motion.
On the other hand, in the structured case
triangles are distributed around a non-zero velocity.
This means that triangles coherently move and have large bulk velocity.
The spread of this region represents a magnitude of the local random velocity.
We attempt to quantitatively separate the motion of a particle into 
the bulk and the local random motions following Salo (1995).
Salo (1995) mainly considered the local random motion 
but did not study the bulk motion.
The bulk motion is also important for understanding the wake structure,
because the motion of particles is largely governed by the bulk motion.
We consider both velocities in detail.
We define the bulk velocity and the local random velocity of particle {\it i}
as the mean velocity averaged from the 10 nearest particles around
particle {\it i} and the difference from this mean velocity
(Use of more particles may include particles that are not coherently moving
while less particles may cause large statistical fluctuation).
These velocities are written as
\begin{equation}
\begin{array}{@{\,}ccl}
\bvs{b}{i} & = & 
        \displaystyle\frac{1}{10} 
	\displaystyle\sum _{\stackrel{j}{{\rm 10 {}nearest}}}{\bm v}_j, \\
\bvs{l}{i} & = & {\bm v}_i - \bvs{b}{i}.
\end{array}
\end{equation}
The (total) velocity dispersion is expressed by using the bulk and 
the local (random) velocity dispersions as
\begin{equation}
\langle v_{\alpha} ^2 \rangle \simeq \langle \vs{b}{\alpha} ^2 \rangle + 
                        \langle \vs{l}{\alpha}^2 \rangle ,
\end{equation}
where $ \alpha $ denotes components of $ x, y, z $-direction.
If there is no correlation of the velocity between neighbor particles,
i.e., particles move randomly,
$ {\bm v} _{{\rm b},i} \simeq 0 $ so that
$ \langle v^2 \rangle ^{1/2} \simeq \langle v_{\rm l} ^2 \rangle ^{1/2} $,
while if there is strong correlation, 
$ |{\bm v} _{{\rm b},i}| \ga |{\bm v} _{{\rm l}, i}| $ so that
$ \langle v^2 \rangle ^{1/2} \simeq \langle v_{\rm b} ^2 \rangle ^{1/2} $.

Figure 11 shows the time evolution of $x$-component of the bulk and 
the local (random) velocity dispersions
in the non-structured and the structured cases in Fig. 4.
The local velocity dispersion is similar in both cases, 
initially the velocity dispersion decreases due to the collision damping 
to attain an equilibrium value with a small magnitude of fluctuation.
On the other hand, the bulk velocity dispersion shows 
a quite different feature.
In the non-structured case, bulk velocity is smaller than local velocity.
This result reconfirms that in the non-structured case, 
particles move randomly but not systematically.
But in the structured case, the bulk velocity grows 
and a large amplitude oscillation starts.
The amplitude is far greater than the local velocity dispersion.
In the structured case, the bulk velocity dominates 
the radial velocity dispersion.
It is clear that the increase and the oscillation in 
the radial velocity dispersion are caused by bulk motion.

The distributions in the local and the bulk velocity spaces are also studied 
by the same method as in Fig. 10, 
in which neighbor particles are marked by the same symbol.
Figure 12 separates the results in Fig. 10b 
into the bulk and the local random velocity components.
The distribution of the local random components shows randomness of motion
as in the non-structured case in Fig. 10a and 
there is no correlation between neighbor particles 
in the local random velocity.
The shape of this distribution does not change with time.
The local random velocity dispersion represents randomness of motion.
The distribution of bulk component in Fig. 12b shows the clockwise revolution 
more clearly than in Fig. 8b.

\newpage

%
%

%
%

\subsection{Characteristic scale of the wake structure 
in self-gravitating particles}

We also studied the characteristic scale of the wake structure.
Number of particles used in Salo (1995) may not be enough to obtain the scale
accurately.
Although Griv (1998) performed local {\it N}-body simulations 
using $N=8000$ self-gravitating particles but without inelastic collision,
he did not measure the scale.
We analyze the characteristic scale of the wake
through simulations with $N>5000$ up to $40000$ particles.

Simulations were done for three different optical 
depth $\tau=0.3$, $0.6$, and $1.2$ with fixed $\epsilon=0.4$ and 
$\hs{r}{h}/2\hs{r}{p}=0.675$.
For the first two simulations, the simulation region is a square
and the width and number of particles are 
$L=230$m ($ L = 170.5\hs{r}{h} $) and $N=5052$ for $\tau=0.3$,
$L=150$m ($ L = 118.6\hs{r}{h} $) and $N=4890$ for $\tau=0.6$.
For $\tau=1.2$, we set the size of the simulation region and 
number of particles
as $L_x = 320$m ($L_x = 240\hs{r}{h}$), $L_y = 350$m ($L_y = 260\hs{r}{h}$), 
and $N = 42780$.
We adopted the autocorrelation analysis to determine the scale.
The autocorrelation of the distribution function of particle position 
on $x$-$y$ plane $n(x,y)$ is defined by
\begin{equation}
\mbox{Corr}(r,s) = \displaystyle\int _{-\infty} ^{\infty} \!dx
	\displaystyle\int _{-\infty} ^{\infty} \!dy \;
	n(x+r, y+s) n(x,y).
\label{eq:corr}
\end{equation}
The distribution function $n(x,y)$ is given by Dirac's $\delta$-function 
$\delta(x)$ as
\begin{equation}
n(x,y) = A \displaystyle\sum_{i} \delta(x-x_i)\delta(y-y_i),
\label{eq:space-dist}
\end{equation}
where $A$ is normalization factor and is chosen as $ A = \sqrt{L_x L_y}/N$
for later convenience.
Substituting Eq.~(\ref{eq:space-dist}) into Eq.~(\ref{eq:corr}),
we obtain
\begin{eqnarray}
\mbox{Corr}(r,s) & = & \displaystyle\frac{L_x L_y}{N^2}
	\displaystyle\sum_{i} \sum _{j} 
	\displaystyle\int _{-\infty} ^{\infty} \!dx \;
		\delta(x+r-x_j)\delta(x-x_i) \nonumber \\ 
	& & {} \times
	\displaystyle\int _{-\infty} ^{\infty} \!dy \;
		\delta(y+s-y_j)\delta(y-y_i).
\label{eq:corr2}
\end{eqnarray}
The integral in terms of $x$ is calculated as 
\begin{equation}
	\displaystyle\int _{-\infty} ^{\infty} \!dx \;
		\delta(x+r-x_j)\delta(x-x_i)
	=\left\{
		\begin{array}{@{\,}lc}
			1 & r=x_j - x_i \\
			0 & \mbox{otherwise}
		\end{array}
	\right. 
\end{equation}
and the same calculation is done for the integral of $y$.
This means that the right hand of Eq.~(\ref{eq:corr2}) except for 
the coefficient represents the number of pairs of particles which satisfy 
the conditions $ r = x_j - x_i $ and $ s = y_j - y_i $, simultaneously.
We take an average of Eq.~(\ref{eq:corr2}) in terms of $r$ and $s$ 
in the range $[r, r+\Delta r]$ and $[s, s+\Delta s]$ as
\begin{eqnarray}
\overline{\mbox{Corr}(r,s)} & = &
	\displaystyle\frac{1}{\Delta r \Delta s} \;
		\int _{r} ^{r+\Delta r} \!dr'
		\int _{s} ^{s+\Delta s} \!ds' \;
		\mbox{Corr}(r',s'),  \nonumber  \\
			& = &
	\displaystyle\frac{L_x L_y}{\Delta r \Delta s N^2} \; n_p(r,s),
\label{eq:corr3}
\end{eqnarray}
where $n_p(r,s)$ is number of pairs of particles 
which satisfy $ r < x_j - x_i < r + \Delta r$ and 
$ s < y_j - y_i < s + \Delta s$.
$ \Delta r $ and $ \Delta s$ are given by using numbers of division of
simulation region $N_x$ and $N_y$ as
$ \Delta r = L_x/N_x $ and $ \Delta s = L_y/N_x $ and 
Eq.~(\ref{eq:corr3}) is expressed as
\begin{equation}
\overline{\mbox{Corr}(r,s)}  = \displaystyle\frac{n_p(r,s)}{N^2/N_x N_y}.
\label{eq:corr4}
\end{equation}
This formula is the same as superposition of particle's distribution 
viewed from each particle.
In the case where particles distribute homogeneously in $x$-$y$ plane,
i.e., in the non-structured case, 
$n_p \sim N^2/N_x N_y$ so that 
$\overline{\mbox{Corr}(r,s)} \sim 1$.

Salo (1995) also adopted the correlation function of 
particle position
by superposition of particle's distribution instead of Fourier transform.
Using this analysis, Salo (1995) displayed the wake-like structure similar to 
gravitational wake in the non-colliding particle's system that was studied by 
Julian and Toomre (1966).
He mainly studied the influence of the boundary conditions on 
the configuration of the wake-like structure.
In the analysis of the autocorrelation function,
we adopt the same method used by Salo (1995).
Figure 13 shows the autocorrelation functions of particle position and 
the particle distribution used in this analysis.
For $\tau=0.3$, inhomogeneity cannot be seen in particle distribution.
The autocorrelation function does not have any clear structure 
without a peak near origin which is caused by 
temporary gravitational binding pointed out by Salo (1995).
On the other hand, as $\tau$ increases, 
a wave-like structure gradually appears.
For $\tau=0.6$ and $1.2$, several clear waves form
in the autocorrelation functions. 
The shape of the waves slightly changes with time 
but the characteristic scale does not change.
Figure 14 shows the cross sections at $y=\mbox{constant}$ 
for $\tau=0.6$ and $1.2$.
The cross sections at different $y$ overlap each other with appropriate shifts.
From these figures, we found that the characteristic wavelengths are
$\hs{\lambda}{ob}/\hs{r}{h} \sim 25$ for $\tau=0.6$ and 
$\hs{\lambda}{ob}/\hs{r}{h} \sim 50$ for $\tau=1.2$, respectively.

Although these waves are the results from
fully developed gravitational perturbation,
we compare our results with the linear perturbation theory.
We confirm that the scale of the wake structure is characterized by
the longest wavelength $\hs{\lambda}{cr}$ that causes
axisymmetric gravitational instability in a thin disk, 
as pointed out by Salo (1995).
In the linear theory,
the most unstable wavelength for a rotating and non-pressure sheet is 
(see, e.g., Binney and Tremaine 1987)
\begin{equation}
\hs{\lambda}{un} = \frac{v_s^2}{G\Sigma},
\label{eq:un}
\end{equation}
where $v_s$ is the random velocity.
When the wake structure is formed, 
\RMS{v_x} no longer represents random velocity
because of the development of coherent motion.
We use the local velocity dispersion $\vsd{l}{x}$ as well as \RMS{v_x} 
for the evaluation of Eq.~(\ref{eq:un}).
On the other hand, 
the wavelength $\hs{\lambda}{cr}$ is given as (Toomre 1964)
\begin{equation}
\hs{\lambda}{cr} = \frac{4\pi^2 G\Sigma}{\kappa^2},
\label{eq:cr}
\end{equation}
where epicyclic frequency $\kappa$ is equal to the angular velocity 
$\Omega_0$ in Saturn's ring.
By using parameters and values of the velocity dispersions obtained by 
our simulations, we evaluate the values of these wavelengths 
in both cases of $\tau=0.6$ and $1.2$ and 
the results are listed in table IV.
The wavelengths $\hs{\lambda}{un}$ calculated from 
the local velocity dispersion are quite different from the results of 
simulations in all cases.
For $\tau=1.2$, $\hs{\lambda}{un}$ calculated from \RMS{v_x} 
is two times lager than observed value and 
the dependence of $\tau$ seems to be different.
On the other hand, $\hs{\lambda}{cr}$ agrees with 
the observed wavelength at all $\tau$.
Thus, the scale of the wake structure is approximately determined by 
$\hs{\lambda}{cr}$ rather than $\hs{\lambda}{un}$.
The result of this analysis reinforces Salo (1992b, 1995). 
For the B-ring, the wavelength calculated from Eq.~(\ref{eq:cr}) with 
B-ring parameter $\Sigma = 120$g/cm$^3$ (e.g., Esposito 1993)
is about $83$m for $\tau=1.0$
(we assumed $\hs{r}{p} = 1$m and $\rho=900$kg/m$^3$, respectively).
This value is consistent with the observational scale $\sim 100$m obtained by 
PPS observation (Esposito et al. 1983a, Esposito 1993).

\newpage

%
%

%
%

\subsection{Velocity change along the line ${\cal Q} \simeq 2$}

Our simulations as well as those in Salo (1995) showed ${\cal Q} \simeq 2$
when the wake structure formed.
The ${\cal Q}$-value is calculated by Eq.~(\ref{eq:ToomreQ}), 
using the equilibrium value of the radial velocity dispersion \RMS{v_x}.
But we showed that the motion of particles includes the large 
systematic motion and \RMS{v_x} no longer represents a magnitude of 
the random motion in the structured case.
By using the local velocity dispersion, 
we obtain ${\cal Q}(\vsd{l}{x}) \simeq 0.38 < 2$ 
so that gravitational instability is very important
(although ring particles do not form persistent clumps due to 
small $\hs{r}{h}/2\hs{r}{p}$ value).

In Fig. 11b, the simulation starts with large random motion 
(large local velocity dispersion)
to avoid gravitational instability. 
When the local velocity dispersion is reduced by 
the collisional dumping to the value for gravitational instability, 
${\cal Q}(\vsd{l}{x}) \la  2$,
the bulk velocity dispersion starts to increase.
The gravitational instability makes particles move coherently and 
the coherent motion dominates the radial velocity dispersion.
The oscillation pattern comes from 
the coherent ``pseudo''-Keplerian bulk motion.
The mean value of the oscillating velocity becomes constant 
independent of time after $5 \Kep$.
Using the mean value of the velocity dispersion,
we obtain ${\cal Q} \simeq 2$.
For simulations with different $\Sigma$ we also obtain ${\cal Q} \simeq 2$.
That is, the mean velocity is larger for larger $\tau$.

However, we have not yet understood 
why ${\cal Q} \simeq 2$ is attained by the systematic motion.
One possibility is gravitational scattering between the transient wakes,
as Salo (1995) suggested.
For the simplicity, we approximate the scattering efficiency by that
between spherical clumps with mass $ \hs{\lambda}{cr}^2 \Sigma$,
which is typical mass of a clump formed by the self-gravitational instability.
Gravitational perturbations pump up velocity dispersion to $\sim r_h \Omega_0$
at one encounter in the case of nearly circular non-inclined orbit
(e.g., Ida 1990).
The ${\cal Q}$-value in this case is (see Eq.~(\ref{eq:ToomreQ}))
\begin{equation}
	\begin{array}{@{\,}ccl}
	{\cal Q} & \sim & 
	  \displaystyle \frac{\Omega_0 \times \hs{r}{h}\Omega_0}{3.36G\Sigma}
		= \frac{\Omega_0 ^2}{3.36 G \Sigma} 
		(\frac{\hs{\lambda}{cr}^2 \Sigma}{3 M_s})^{1/3}, \\
		 &      & {} \simeq 2.4. 
	\end{array}
\end{equation}
Note that the dependences of ${\cal Q}$ on particle mass,
$\Sigma$, $\Omega_0$, and $M_s$ are canceled by $\hs{\lambda}{cr}$
(see Eq.~(\ref{eq:cr}))
and ${\cal Q}$ takes a constant value $ \sim 2$.
When $\Sigma$ is larger, more massive clumps forms and 
velocity dispersion is increased by the stronger gravity.
This result is consistent with the simulation results, however, 
more detail and more correct treatment is needed for the scattering between 
the wakes.

\newpage

%
%
%
%

\section{Conclusion and Discussion}

In this study,
we performed local $N$-body simulations including 
the mutual gravitational force between ring particles as well as 
the direct (inelastic) collision with identical (up to $N \simeq 40000$) 
particles.
Salo (1995) showed that the spatial structures 
(wake and clumps) arise spontaneously
in a dense, self-gravitating, and collisionally dumping particle system 
due to gravitational instability.
He also showed that as the wake forms,
the radial velocity dispersion increases with the relation ${\cal Q} \sim 2$,
where ${\cal Q}$ is Toomre's non-dimensional parameter.
We reconfirmed his results (section 3.1).
Furthermore, our simulations showed a regular oscillatory pattern in 
the velocity dispersion which was not found in Salo (1995).
We analyzed the motion of particles in the wake structure in detail 
and clarified that the intrinsic physics of the regular oscillation is
coherent ``pseudo''-Keplerian motion induced by self-gravitational instability
(section 3.2).

The wake-like structure is not steady but changes with time, 
continuously forming clumps due to 
gravitational instability and dissolving due to shear motion,
which looks like the turbulent fluid.
We clearly showed that in such circumstances,
particles no longer move randomly but coherently. 
The coherent motion results in transient Keplerian motion 
with similar orbital elements of the particles in a clump, 
presumably because of mutual forcing by self-gravity.
Such a forcing mechanism is proposed to explain the elliptic ring around Uranus
(Goldreich and Tremaine 1979).
If we separate the particle motion into the bulk and the local random motions 
following Salo (1995), 
in the structured system, the motion is dominated by the bulk motion and
the local random motion is small with almost steady distribution.
Owing to the coherent motion, the radial velocity dispersion oscillates 
in a period of the order of the Keplerian period, 
but the period observed in our simulations is somewhat larger than 
one Keplerian period, 
which may be because of friction through inelastic collisions.

We obtained the wavelength of the wake through autocorrelation analysis
in section 3.3.
The wavelength observed in our simulation is approximately given
by the longest wavelength $\hs{\lambda}{cr}$ in the linear theory of 
axisymmetric gravitational instability in a thin disk. 
The wavelength $\hs{\lambda}{cr}$ is given by 
$\hs{\lambda}{cr} = 4\pi^2G\Sigma/\kappa^2$ (Toomre 1966).
The gravitational scattering between the clumps with typical mass
$\sim \Sigma \hs{\lambda}{cr} ^2$ may be responsible for ${\cal Q} \sim 2$
as shown in section 3.4, however,
we need more detailed analysis on non-linear effects of 
the strong wake structure.

The structure observed in our and Salo's simulations is non-axisymmetric one.
The observation in Saturn's A-ring produced the evidence of 
the asymmetric structure which may be caused by the wake
(Dones et al. 1993).
The characteristic scale of the structure observed in our simulations 
corresponds to the size of an order of $\sim 100$m for B-ring's parameters, 
which is consistent with the scale of the microstructure observed 
by PPS observation.
The microstructure may correspond to the wake structure we found.
More information about structure of the ring will be obtained 
by Cassini mission.

Our simulation results could be applicable to other optically thick 
particulate disks in which 
both the mutual gravitational force between particles and 
the collisional dumping are important.
One of such systems is the Uranian elliptic ring system.
Uranian rings are also optically thick system (e.g., Esposito 1993)
and a wake-like structure may arise due to gravitational instability.
In the framework of a giant impact model of the origin of moon,
it is considered that an optically thick debris disk formed around 
proto-earth after the impact 
and that the moon would have accreted in the disk
(Hartmann and Davis 1975, Cameron and Ward 1976).
According to $N$-body simulations performed by Ida et al. (1997)
and Kokubo et al. (1999),
spiral patterns develop in the early stage of the simulation and 
it results in rapid transfer of angular momentum and mass.

In the next paper, we will address 
angular momentum transfer enhanced by gravitational torque of 
the wake structure in detail.

\newpage

%
%

\begin{center}
{\Large{\bf ACKNOWLEDGMENTS}}
\end{center}

The authors are indebted to S. Araki and K. Ohtsuki for
their variable comments and useful suggestions.
We thank H. Tanaka, H. Emori ,and K. Nakazawa 
for continuous encouragement.
We also thank J. Makino and E. Kokubo for technical advice about HARP-2.
This work was supported by the Grant-in-Aid of the Japanese Ministry of
Education, Science, Sports, and Culture (09440089).

\newpage

%

%
%

\begin{center}
{\Large{\bf REFERENCES}}
\end{center}

\refi{Araki, S. and S. Tremaine}{1986}
{The Dynamics of Dense Particle Disks}
{Icarus}{65}{83-109}

\refi
{Araki, S.}{1988}
{The Dynamics of Particles Disks. II. Effects of Spin Degrees of Freedom}
{Icarus}{65}{83-109}

\refi
{Araki, S.}{1991}
{The Dynamics of Particles Disks. III. Dense and Spinning Particle Disks}
{Icarus}{90}{139-171}

\refb
{Binney, J. and S. Tremaine}{1987}
{Galactic Dynamics}{Princeton Univ. Press, Princeton. NJ}{283pp.}

\refi
{Bridges, F. G., A. Hatzes, and D. N. C. Lin}{1984}
{Structure, stability and evolution of Saturn's rings}
{Nature}{309}{333-335}

\refi
{Cameron, A. G. W. and W. R. Ward}{1976}
{The origin of the Moon}
{Proc. Lunar Planet Sci. Conf.}{7}{120-122}

\refi
{Dilley, J. P.}{1993}
{Energy Loss in Collisions of Icy Spheres:
Loss Mechanism and Size-Mass Dependence}
{Icarus}{105}{225-234}

\refi
{Dones, L., J. N. Cuzzi, and M. R. Showalter}{1993}
{Voyager Photometry of Saturn's A Ring}
{Icarus} {105}{184-215}

\refi
{Esposito, L. W., M.O'Callaghan, and R. A. West}{1983a}
{The Structure of Saturn's Rings: Implications from the 
Voyager Stellar Occultation}
{Icarus} {56} {439-452}

\refi
{Esposito, L. W., M.O'Callaghan, K. E. Simmons, C. W. Hord, R. A. West, 
A. L. Lane, R. B. Pomphery, D. L. Coffeen, and M. Sato}{1983b}
{Voyager Photopolarimeter Stellar Occultation of Saturn's Rings}
{J. Geophys. Res.} {88} {8643-8649}

\refi
{Esposito, L. W.} {1993}
{Understanding planetary rings}
{Annu. Rev. Earth Planet. Sci} {21} {487-523}

\refi
{Goldreich, P. and S. Tremaine} {1978}
{The Velocity Dispersion in Saturn's Rings}
{Icarus} {34} {227-239}

\refi
{Goldreich, P. and S. Tremaine}{1979}
{Precession of the $\epsilon$ ring of Uranus}
{Astron. J.} {84} {1638-1641}

\refi
{Goldreich, P. and S. Tremaine} {1982}
{The dynamics of planetary rings}
{Ann. Rev. Astron. Astrophys.} {20} {249-283}

\refi
{Griv, E.}{1998}
{Local stability criterion for the Saturnian ring system}
{Planet. Space Sci.} {46} {615-628}

\refi
{Hartmann, W. K. and D. R. Davis} {1975}
{Satellite-Sized Planetesimals and Lunar Origin}
{Icarus} {24} {504-515}

\refi
{Hatzes, A. P., F. G. Bridges, and D. N. C. Lin} {1988}
{Collision properties of ice spheres at low impact velocities}
{Mon. Not. R. Astron. Soc.} {231} {1091-1115}

\refi
{Hill, G. W.} {1878}
{Researches in the Lunar Theory}
{Amer. J. Math.} {1} {5-26, 129-147, 245-260}

\refi
{Ida, S} {1990}
{Stirring and Dynamical Friction Rates of Planetesimals in the Solar
Gravitational Field}
{Icarus} {88} {129-145}

\refi
{Ida, S. and J. Makino} {1992}
{{\it N}-body Simulation of Gravitational Interaction between 
Planetesimals and a Protoplanet
I. Velocity Distribution of Planetesimals}
{Icarus} {96} {107-120}

\refi
{Ida, S., R. M. Canup, and G. R. Stewart} {1997}
{Lunar accretion from an impact-generated disk}
{Nature} {389} {353-357}

\refi
{Julian, W. H. and A. Toomre} {1966}
{Non-axisymmetric responses of differentially rotating disks of starts}
{Astrophys. J.} {146} {810-827}

\refpre
{Kokubo, E., J. Makino, and S. Ida} 
{in preparation}

\refi
{Lin, D. N. C. and P. Bodenheimer} {1981}
{On the stability of Saturn's rings}
{Astrophys. J. letter} {248} {L83-L86}

\refi
{Lukkari, J.} {1981}
{Collisional amplification of density fluctuations in Saturn's rings}
{Nature} {292} {433-435}

\refi
{Makino, J. and S. J. Aarseth} {1992}
{On a Hermite integrator with Ahmad-Cohen Scheme for Gravitational 
Many-Body Problems}
{Publ. Astron. Soc. Japan.} {44} {141-151}

\refi
{Makino, J., E. Kokubo, and M. Taiji} {1993}
{HARP: A Special-Purpose Computer for {\it N}-Body Problem}
{Publ. Astron. Soc. Japan.} {45} {349-360}

\refi
{Makino, J., M. Taiji, T. Ebisuzaki, and D. Sugimoto} {1997}
{GRAPE-4: A Massively Parallel Special-Purpose Computer for Collisional 
{\it N}-body simulations}
{Astrophys. J.} {480} {432-446}

\refi
{Nakazawa, K. and S. Ida} {1988}
{Hill's Approximation in the Three-Body Problem}
{Prog. Theor. Phys. Suppl.} {96} {167-174}

\refi
{Ohtsuki, K.} {1992}
{Equilibrium Velocities in Planetary Rings with Low Optical Depth}
{Icarus} {95} {265-282}

\refi
{Ohtsuki, K.} {1993}
{Capture Probability of Colliding Planetesimals:
Dynamical Constraints on Accretion of Planets, Satellites, and Ring Particles}
{Icarus} {106} {228-246}

\refi
{Ohtsuki, K.} {1999}
{Evolution of Particle Velocity Dispersion in a Circumplanetary Disk 
Due to Inelastic Collisions and Gravitational Interactions}
{Icarus} {137} {152-177}

\refb
{Press, W. H., B. P. Flannery, S. A. Teukolsky, and W. T. Vetterling}{1986}
{Numerical Recipes}
{Cambridge Univ. Press, London/NewYork}{472pp.}

\refi
{Richardson, D. C.} {1994}
{Tree code simulations of planetary rings}
{Mon. Not. R. Astron. Soc.} {269} {493-511}

\refi
{Salo, H.} {1991}
{Numerical Simulations of Dense Collisional Systems}
{Icarus} {90} {254-270}

\refi
{Salo, H.} {1992a}
{Numerical Simulations of Dense Collisional Systems. 
II. Extended Distribution of Particle Sizes}
{Icarus} {96} {85-106}

\refi
{Salo, H.} {1992b}
{Gravitational wakes in Saturn's rings}
{Nature} {395} {619-621}

\refi
{Salo, H.} {1995}
{Simulations of Dense Planetary Rings. 
III. Self-Gravitating Identical Particles}
{Icarus} {117} {287-312}

\refi
{Smith, B. A., L. Soderblom, R. Batson, P. Bridges, J. Inge, 
H. Masursky, E. Shoemaker,
R. Beebe,
J. Boyce, G. Briggs,
A. Bunker, S. A. Collins, C. J. Hansen, T. V. Johnson, J. L. Mitchell,
R. J. Terrile, 
A. F. Cook II,
J. Cuzzi, J. B. Pollack, 
G. E. Danielson, A. P. Ingersoll, 
M. E. Davies, 
G. E. Hunt,
D. Morrison, 
T. Owen,
C. Sagan, J. Veverka,
R. Strom, 
and
V. E. Suomi} {1982}
{A New Look at Saturn System: The Voyager 2 Images}
{Science} {215} {504-537}

\refi
{Sugimoto, D., Y. Chikada, J. Makino, T. Ito, T. Ebisuzaki, and M. Umemura} 
{1990}
{A special-purpose computer for gravitational many-body problem}
{Nature} {345} {33-35}

\refi
{Supulver, K. D., F. G. Bridges, and D. N. C. Lin} {1995}
{The Coefficient of Restitution of Ice Particles in Glancing Collisions: 
Experimental Results for Unfrosted Surfaces}
{Icarus} {113} {188-199}

\refi
{Toomre, A.} {1964}
{On the gravitational stability of a disk of stars}
{Astrophys. J.} {139} {1217-1238}

\refi
{Ward, W. R.} {1981}
{On the radial structure of Saturn's ring}
{Geophys. Res. Let.} {8} {641-643}

\refi
{Wisdom, J. and S. Tremaine} {1988}
{Local simulations of planetary rings}
{Astron. J.} {95} {925-940}

\newpage

%
%

\begin{center}
\Large{\bf TABLE}
\end{center}

\begin{center}
Table I. 
Equilibrium radial velocity dispersions
in simulations with parameters used by Salo(1995);
simulations with various optical depth $\tau$ for $\epsilon=0.5$,
$\hs{r}{h}/2\hs{r}{p}=0.82$.
\vspace{0.5cm}

\begin{tabular}{ccccc}
\hline
       &     &         & \multicolumn{2}{c}{\RMSn{v_x}} \\ \cline{4-5}
$\tau$ & $N$ & $L/\hs{\lambda}{cr}$ & Our results & Salo(1995)       \\
\hline
0.1  & 400  & 13.5 & $2.094 \pm 0.084$ & $2.05 \pm 0.08$  \\
0.2  & 800  & 6.7  & $2.157 \pm 0.084$ & $2.16 \pm 0.08$  \\
0.3  & 1200 & 4.5  & $2.443 \pm 0.149$ & $2.46 \pm 0.13$ \\
0.4  & 1600 & 3.4  & $2.906 \pm 0.261$ & $3.01 \pm 0.27$  \\
0.5  & 2000 & 2.7  & $3.987 \pm 0.663$ & $4.05 \pm 0.53$  \\
0.6  & 2400 & 2.2  & $5.094 \pm 1.137$ & $5.84 \pm 1.42$  \\ \hline
\end{tabular}
\end{center}

\vspace{1.0cm}

\begin{center}
Table II.
Simulations with various velocity-independent restitution coefficient 
$\epsilon$ for $\tau=0.4$ and $\hs{r}{h}/2\hs{r}{p}=0.82$.
\vspace{0.5cm}

\begin{tabular}{ccc}
\hline
           & \multicolumn{2}{c}{\RMSn{v_x}} \\ \cline{2-3}
$\epsilon$ & Our results & Salo(1995)       \\
\hline
0.1  &  $4.500 \pm 0.589$ & $ 4.39 \pm 0.671$  \\
0.2  &  $4.578 \pm 0.762$ & $ 4.27 \pm 0.732$  \\
0.3  &  $4.221 \pm 0.741$ & $ 3.84 \pm 0.671$  \\
0.4  &  $4.059 \pm 0.589$ & $ 3.48 \pm 0.427$  \\
0.5  &  $2.906 \pm 0.261$ & $ 3.05 \pm 0.427$  \\
0.6  &  $2.584 \pm 0.168$ & $ 2.61 \pm 0.183$  \\
0.7  &  $3.670 \pm 0.114$ & $ 3.60 \pm 0.073$  \\ \hline
\end{tabular}
\end{center}

\vspace{1.0cm}

\begin{center}
Table III.
Simulations with various number of particles for $\tau=1.2$, $\epsilon=0.4$,
and $\hs{r}{h}/2\hs{r}{p}=0.675$.
\vspace{0.5cm}

\begin{tabular}{ccccccc}
\hline
$N$   & $L_x/\hs{r}{h}$ & $L_y/\hs{r}{h}$ & $L_x/\hs{\lambda}{cr}$ & 
$L_y/\hs{\lambda}{cr}$ & $v_x/\hs{r}{h}\Omega_0$ & err \\ \hline
2460  & 60  & 60  & 1.45 & 1.45 & 5.45 & 1.61  \\
5054  & 85  & 85  & 2.07 & 2.07 & 5.28 & 1.49  \\
10696 & 60  & 260 & 1.45 & 6.27 & 6.12 & 1.51  \\
21290 & 120 & 260 & 2.90 & 6.27 & 7.16 & 1.38  \\
42780 & 240 & 260 & 5.80 & 6.27 & 6.55 & 0.712 \\ \hline
\end{tabular}
\end{center}

\vspace{1.0cm}

\begin{center}
Table IV.
Characteristic wavelength for large computational simulations.
\vspace{0.5cm}

\begin{tabular}{cccccccccc}
\hline
$\tau$ & 
\ha{ob} & \ha{cr} & \he{un}{\RMSm{v_x}} &
\he{un}{\vsd{l}{x}} &  \RMSn{v_x} & $\vsdn{l}{x}$ \\ \hline 
0.3 &  -       & $10.31$ & $22.67$ & $15.56$ & 
        $1.72 \pm 0.0267$ & $1.42 \pm 0.014 $ \\
0.6 & $\sim25$ & $20.61$ & $19.88$ & $5.51$  & 
        $2.27 \pm 0.130$  & $1.20 \pm 0.037 $ \\
1.2 & $\sim50$ & $41.22$ & $82.09$ & $3.75$  & 
        $6.54 \pm 0.712$  & $1.40^*           $ \\
\hline
\end{tabular}
\end{center}
$^*$ This value is not time-averaged.

\newpage

%
%

\begin{center}
\Large{\bf FIGURE CAPTION}
\end{center}

\re
{\bf FIG. 1.}
Schematic illustration of computational domain.
Original domain is a box with thick line.
It is surrounded by eight copied domains.
The shear velocities of the outside and inside boxes are
 $ - 3 \Omega_0 L_x/2 $ and
$ + 3\Omega_0 L_x/2 $.
Computational area is divided into nine subareas (broken lines) and 
we make a virtual area for each subregion in which the subarea is centered.
We calculate gravitational forces in the virtual area.
For example, consider the particle denoted by a cross in the subregion
(represented by dark shaded region).
This subregion has virtual region represented by light shaded region.

\re
{\bf FIG. 2a.}
Snapshots of the particle distribution with 
various optical depth ($\tau = 0.1$ to $0.6$).
Upper and lower panels are face-on and edge-on distribution.
In all cases, the restitution coefficient $\epsilon$ and 
ratio $\hs{r}{h}/2\hs{r}{p}$ are 
$\epsilon = 0.5$ and $\hs{r}{h}/2\hs{r}{p} = 0.82$, respectively.
The width of the simulation region is $L=112$m ($L = 68.3 \hs{r}{h}$).

\re
{\bf FIG. 2b.}
The same as Fig. 2a but simulations with various 
restitution coefficient ($\epsilon = 0.5, 0.6$, and $0.7$) for $\tau = 0.4$.

\re
{\bf FIG. 2c.}
The same as Fig. 2a and 2b but simulations with various 
$\hs{r}{h}/2\hs{r}{p}$ for
$\tau = 0.4$ and $\epsilon = 0.5$.
The width is set as $L = 59.3 \hs{r}{h}$ ($L=80$m)
for $\hs{r}{h}/2\hs{r}{p}=0.675$, 
$L = 68.3 \hs{r}{h}$ ($L=112$m) for $\hs{r}{h}/2\hs{r}{p}=0.82$ , and
$L = 70.0 \hs{r}{h}$ ($L=140$m) for $\hs{r}{h}/2\hs{r}{p}=1.00$ , respectively.

\re
{\bf FIG. 3.}
Equilibrium values of the velocity dispersion as a function of
the optical depth $\tau$ for $\epsilon = 0.5$ (a) and 
the restitution coefficient $\epsilon$ for $\tau=0.4$ and $\tau=0.1$ (b).
Our results and Salo(1995) are represented by filled circles and 
open triangles, respectively.
For our results, both \RMS{v_x} and \RMS{v_z} are plotted in (a) 
but only \RMS{v_x} in (b).
For Salo(1995), only \RMS{v_x} is plotted in both (a) and (b).
Surface escape velocity is $\hs{v}{es} = \sqrt{2Gm/\hs{r}{p}}$,
where $m$ and $\hs{r}{p}$ are mass and radius of a particle, respectively.
Both dotted and broken lines in (a) are calculated from Toomre's 
${\cal Q}$-value (Eq.~(\ref{eq:ToomreQ}))
as being ${\cal Q}=1$ and ${\cal Q}=2$.
All velocities are scaled by $\hs{r}{h} \Omega_0$.

\re
{\bf FIG. 4.}
Time evolution of the radial velocity dispersions for $\tau=0.3$ (broken line) 
and $\tau=1.2$  (solid line), 
which correspond to the non-structured and the structured cases, respectively.
In both simulations, $\epsilon$ and $\hs{r}{h}/2\hs{r}{p}$ are 
$\epsilon = 0.4$ and $\hs{r}{h}/2\hs{r}{p} = 0.675$.
The number of particles is $N = 5052$ in both cases.
Typical spatial distributions of both cases are superimposed.
For $\tau = 0.3$, edge of the distribution is trimmed away and 
the distribution displays the same area as that for $\tau = 1.2$.
The width of each box is $115$m ($85.2\hs{r}{h}$).

\re
{\bf FIG. 5.}
Dependence of the velocity dispersion on 
number of particles (equivalently, area of the computational region).
In each simulation, $\tau = 1.2$, $\epsilon = 0.4$, 
and $\hs{r}{h}/2\hs{r}{p} = 0.675$.

\re
{\bf FIG. 6.}
Spectrum analysis of a period of oscillation observed in 
the evolution curve of \RMS{v_x} in the non-structured (broken line) and 
the structured cases (solid line).

\re
{\bf FIG. 7.}
Influence of simulation parameters on a period of oscillation observed in 
the evolution curve of \RMS{v_x}.
(a) Various numbers of particles but the same parameters of
$\tau = 1.2$, $\epsilon = 0.4$, and $\hs{r}{h}/2\hs{r}{p} = 0.675$.
(b) Various optical depth but the same parameters of
$\epsilon = 0.4$, and $\hs{r}{h}/2\hs{r}{p} = 0.675$.
Simulations are done with the same simulation region as 
$L = 59.3\hs{r}{h}$ ($L=80$m) and optical depth is controlled 
by changing number of particles.
Used particles in each simulation are 
$N = 1630$ for $\tau = 0.8 $,
$N = 2038$ for $\tau = 1.0 $, and
$N = 2444$ for $\tau = 1.2 $, respectively.
(c) Various restitution coefficients but the same parameters of
$\tau = 1.1$ and $\hs{r}{h}/2\hs{r}{p} = 0.675$.
Simulation region is the same as the case of (b), but
number of particles is $N=2240$.
(d) Various $\hs{r}{h}/2\hs{r}{p}$ but the same parameters of
$\tau = 0.6$ and $\epsilon = 0.5$.
Width of simulation region and number of particles are
$L = 68.2\hs{r}{h} $ ($L=112$m) and $N = 2400$ 
for $\hs{r}{h}/2\hs{r}{p} = 0.82$ and
$L = 70.0\hs{r}{h} $ ($L=140$m) and $N = 3744$ 
for $\hs{r}{h}/2\hs{r}{p} = 1.00$,
respectively.

\re
{\bf FIG. 8a.}
Time sequence of spatial distribution of the structured case ($\tau = 1.2$) 
from $t = 8.91 \Kep$ to $t = 10.12 \Kep$.
The width of region and number of particles are $L=115$m
($L = 85.2\hs{r}{h}$) and $N=5052$, respectively.

\re
{\bf FIG. 8b.}
Time sequence of the corresponding velocity distribution to Fig. 8a.
For each particle, the local shear velocity ($0$,$3 \Omega _0 x_i/2$,$0$) is 
subtracted.
Arrows represent the direction of major long axis of distribution
at each time.

\re
{\bf FIG. 8c.}
Cumulative distribution of the radial velocity $v_x$ in Fig. 8b.
Broken line with open circle, dashed-dotted line with open triangle, and
dotted line with cross denote distributions at $t=9.55 \Kep$, $t=9.71 \Kep$,
and $t=9.87 \Kep$, respectively.
Solid line denotes cumulative Gaussian distribution
$ P(s) =  (1 + \mbox{erf}(s/\sqrt{2}))/2 $, where $\mbox{erf}(s)$ is
the error function.
The distributions are scaled by the velocity dispersions at each time.
The instantaneous velocity dispersion is 
$\langle v_x ^2 \rangle ^{1/2} _{ins}/\hs{r}{h} \Omega_0 = 6.100$ 
at $t=9.55 \Kep$,
$\langle v_x ^2 \rangle ^{1/2} _{ins}/\hs{r}{h} \Omega_0 = 7.115$ 
at $t=9.71 \Kep$,
and 
$\langle v_x ^2 \rangle ^{1/2} _{ins}/\hs{r}{h} \Omega_0 = 4.033$ 
at $t=9.87 \Kep$, respectively.
The time averaged velocity dispersion is 
$\langle v_x ^2 \rangle ^{1/2} _{ave}/\hs{r}{h} \Omega_0 = 5.862$.
The cumulative number is scaled by the total number of particles.

\re
{\bf Fig. 8d.}
Time evolutions of $x$-component of position (left) and velocity (right)
for certain three particles in the structured case ($\tau = 1.2$).
Solid, broken, and line-dotted lines illustrate the evolution curves
of particles.
Dotted lines denote positions of boundary of $x$-direction, 
the range of simulation domain is $[-42.62 \hs{r}{h}, 42.62 \hs{r}{h}]$.
In this figure, curves are connected continuously by taking into account
boundary conditions.
But in fact, evolution curve for a particle over a boundary discontinues
there and 
the curve corresponding to incoming particle starts from the opposite side.

\re
{\bf Fig. 8e.}
Time evolutions of $v_x$ for particles used in Fig. 8d.
Each origin moves with appropriate shift and each dotted line denotes
the line of $v_x = 0$.

\re
{\bf FIG. 9.}
Velocity distributions of the non-structured case ($\tau = 0.3$).
(a) Particle distribution in velocity space at $t=14.3 \Kep$.
(b) Cumulative distribution of radial velocity scaled by \RMS{v_x}.
The velocity dispersion is 
$\langle v_x ^2 \rangle ^{1/2}/\hs{r}{h} \Omega_0 = 1.720$.
Broken line with open circle, dashed-dotted line with open triangle, and
dotted line with cross denote distributions at $t=11.1 \Kep$, $t=12.7 \Kep$,
and $t=14.3 \Kep$, respectively.
Solid line denotes cumulative Gaussian distribution.

\re
{\bf FIG. 10.}
The relation of marked particles in the real and the velocity space
in the non-structured case ($\tau=0.3$) (a) and 
structured case ($\tau=1.2$) (b).
Particles within certain three regions with the radius $5\hs{r}{h}$ 
which are located at ($-34\hs{r}{h}$,$11\hs{r}{h}$), 
($10\hs{r}{h}$,$30\hs{r}{h}$), and ($23\hs{r}{h}$,$3\hs{r}{h}$) 
in the real space are marked by
symbols as filled triangle, circle, and square, respectively.
The other particles are represented by small dots for convenience.
In the non-structured case, edge of the spatial distribution is also trimmed.

\re
{\bf FIG. 11.}
Evolution curves of the bulk and local velocity dispersion
in the non-structured case (a) and the structured case (b).
Dotted line ,dashed line with filled triangle, and solid line denote 
local, bulk velocity dispersion, and \RMS{v_x}, respectively.

\re
{\bf FIG. 12.}
The Same as Fig. 10b but for particle distributions 
in the local and bulk velocity space.
(a) Distribution for local velocity.
(b) Distribution for bulk velocity.
The local shear velocity of each particle is also eliminated.

\re
{\bf FIG. 13.}
Autocorrelation functions of particle positions 
in the case of $\tau=0.3$ (a), $0.6$ (b), and $1.2$ (c).
In all simulations, $\epsilon = 0.4$ and $\hs{r}{h}/2\hs{r}{p}=0.675$ 
are set up.
We also showed the original snapshots used in this analysis.

\re
{\bf FIG. 14.}
Cross sections of autocorrelation functions for  $\tau=0.6$ (a) and 
$\tau=1.2$ (b) in Fig.~13.
Each cross section is shifted in the $x$-direction for peak positions 
to coincide with each other.

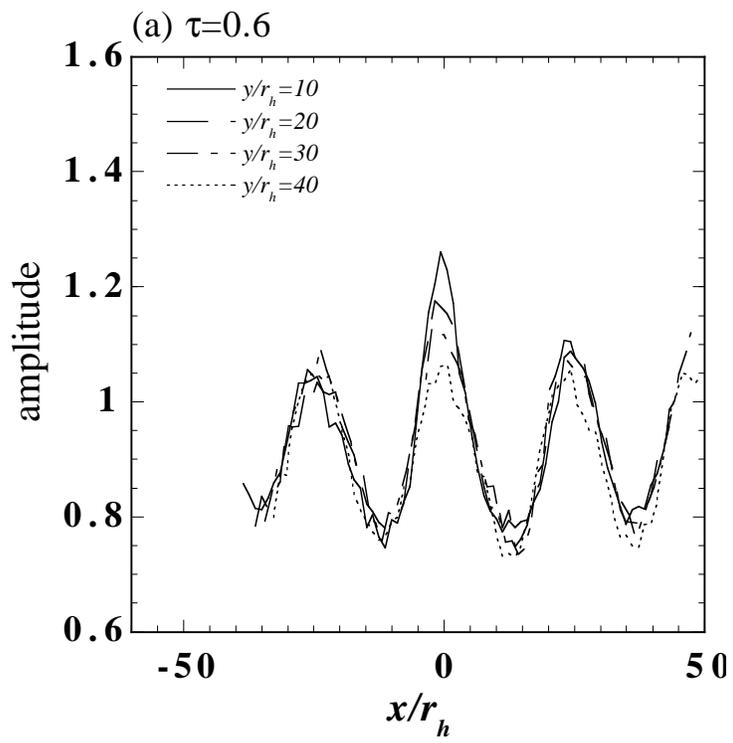
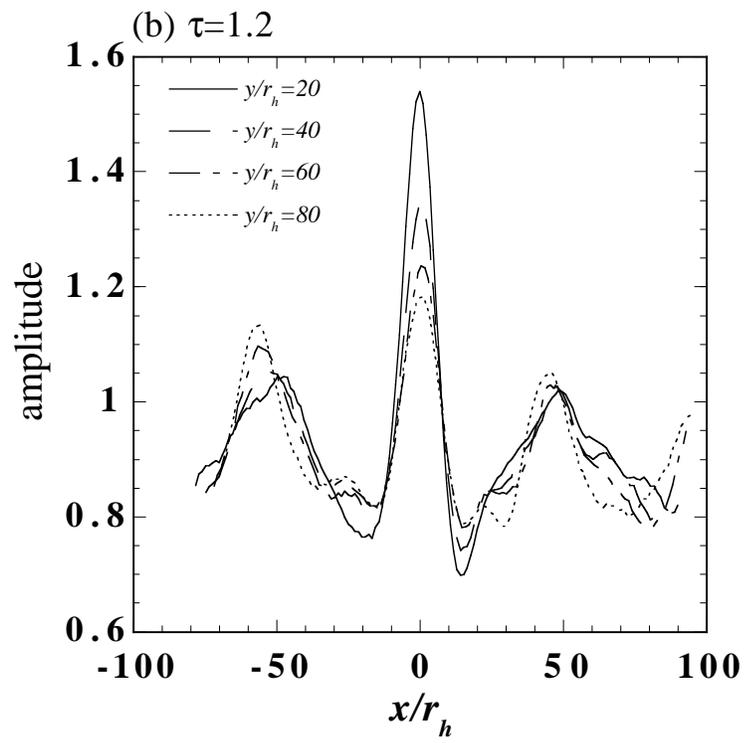
\end{document}